\documentstyle[12pt,epsfig]{article}
\begin{document}
\scriptsize Journal of The Korean Astronomical Society 44: 1 $\sim$ 11, 2011~~~~~~~~~~~~~~~~~~doi:10.5303/JKAS.2011.44.1.1
\bigskip
\bigskip
\bigskip

\begin{center}
{\large \bf A CATALOG OF 120 NGC OPEN STAR CLUSTERS}\\
\bigskip
\normalsize {\bf A. L. Tadross}\\
\bigskip
\scriptsize National Research Institute of Astronomy \& Geophysics, Helwan, Cairo, Egypt.\\
email: altadross@yahoo.com
\end{center}
\bigskip

\begin{abstract}
A sample of 145 JHK--2MASS observations of NGC open star clusters is studied, of which 132 have never been studied before. Twelve are classified as non-open clusters and 13 are re-estimated self-consistently, after applying the same methods in order to compare and calibrate our reduction procedures. The fundamental and structural parameters of the 120 new open clusters studied here are derived using color-magnitude diagrams of JHK Near-IR photometry with the fitting of solar metallicity isochrones. We provide here, for the first time, a catalog of the main parameters for these 120 open clusters, namely, diameter, distance, reddening and age.
\end{abstract}

\bigskip

\section{INTRODUCTION}

Systematic studies of open star clusters (OCs) are very important for understanding galactic structure and star formation processes, as well as stellar evolution and evolution of the Galactic disk. By utilizing color-magnitude diagrams (CMDs) of the stars observed in the near-infrared (NIR) bands, it is possible to determine the properties of open clusters such as distance, reddening, age and metallicity. Such parameters are necessary for studying clusters and Galactic disk. The Galactic (radial and vertical) abundance gradient also can be studied using OCs (Hou, Prantzos, \& Boissier 2000; Chen, Hou, \& Wang 2003; Kim \& Sung 2003; Tadross 2003; Kim et al. 2005). According to some estimations, there are as many as 100,000 OCs in our Galaxy, but less than 2000 of them have been discovered and cataloged (Piskunov et al. 2006 and Glushkova et al. 2007). Actually, not all the clusters discovered so far have their basic photometrical parameters available in the current literature. However, such as yet unstudied ones are, in general, poorly populated and/or distant and/or rejected against dense foreground/background fields. In all these cases, field-star contamination is a fundamental issue that has to be dealt with before robust astrophysical parameters are derived. So, our aim in the present paper, a continuation of a series of papers, is to determine the main astrophysical properties of previously unstudied OCs using modern databases (Tadross 2008a, 2008b \& Tadross 2009a, 2009b and references therein).
In this respect, the present study introduces the first photometric analysis of CMDs of a sample of 120 OCs. Note that, the JHK--2MASS system gives more accurate fitting, especially for the lower portions, so that the turnoff points, distance modulii, ages and photometrical membership can be improved.

The JHK Near-IR 2MASS catalog has the advantage of being a homogeneous database, enabling us to observe young clusters in their dusty environments and reach their outer regions where low mass stars dominate. For the task of calibration, the main parameters of 13 NGC previously studied clusters are re-estimated and compared with those available in the literature. The clusters were arbitrarily selected from among the unstudied NGC candidates. The only previously known information about the investigated clusters (132 OCs) are the coordinates and the apparent diameters, which were obtained from the WEBDA\footnote{\it http://obswww.unige.ch/webda} site and the last updated version of DIAS\footnote{\it http://www.astro.iag.usp.br/$\sim$wilton/} collection (version 3.0, 2010 April 30). These are sorted by right ascensions and listed in Table 1 in the Appendix. The quality of the data has been taken into account and the properties of all the clusters have been estimated by applying the same method to each of them.

This paper is organized as follows. The data extraction and preparation are described in Section 2.
Center determination and radial density profile are clarified in Sections 3 and 4 respectively.
Sections 5, 6 and 7 are devoted to field-star decontamination, analysis of CMDs and Galactic geometric
distances, respectively. The final conclusions are stated in Section. 8.
All tables are reported in the Appendix.
%--------------------------------------------------------------------------------------
\section{DATA EXTRACTION AND PREPARATION}
The Two Micron All Sky Survey (2MASS) of Skrutskie et al. (2006) collected 25.4 T-bytes of raw imaging data covering 99.998\% of the celestial sphere. Observations were conducted using two dedicated 1.3 m diameter telescopes located at Mount Hopkins, Arizona, and Cerro Tololo, Chile. $256 \times 256$ NICMOS3 (HgCdTe) arrays manufactured by Rockwell International Science Center (now Rockwell Scientific), were used which give field-of-view of $8.'5 \times 8.'5$ and pixel scale of $2''$ pixel$^{-1}$. The photometric system comprise $J$ (1.25 $\mu$m), $H$ (1.65 $\mu$m) and $K_S$ (2.16 $\mu$m) bands, where  the ``$K$-short'' ($K_S$) filter excludes wavelengths longward of 2.31 $\mu$m to reduce thermal background and airglow and includes wavelengths as short as 2.00 $\mu$m to maximize bandwidth.

Data extraction has been performed using the known tool VizieR for 2MASS\footnote{\it http://vizier.u-strasbg.fr/viz-bin/VizieR?-source=2MASS} database. The investigated clusters have been selected from WEBDA and DIAS catalogues. The clusters data have been extracted at a preliminary radius of about 2 times larger than the values provided by WEBDA or/and DIAS, in circular areas centered at the coordinates of the clusters centers, covering possibly the clusters' coronas (see Section 4). Note that, the field stars, mostly on the disk, contaminate the CMDs of low-latitude clusters, particularly at faint magnitudes and red colors. Therefore, suitable clusters should have enough members with prominent sequences in their CMDs. Further, the clusters should have good blue images on the Digitized Sky Surveys\footnote {\it http://cadcwww.dao.nrc.ca/cadcbin/getdss} clearly separated from the background field (see the example of NGC 7801 image in Fig. 1).

A cutoff of photometric completeness limit at $J<16.5$ mag is applied on the photometric data to avoid over-sampling (Bonatto et al. 2004).
To retain the brightest stars of the cluster, the data extraction was restricted to faint stars with
errors in {\it J, H} and {\it $K_S$} smaller than 0.2\,mag. In order to establish a clean CMD for each
cluster, its control field should be compared and color -- magnitude filters should be applied to the
cluster and its field sequences. Membership criteria is adopted for the location of the stars in the
clean CMD, where the stars located away from the main sequences are excluded (Bonatto et al. 2005).
The maximum departure accepted here is about 0.10 - 0.15 mag (see the example in Fig. 2).
%---------------------------------------------------------------------
\begin{figure}
\centering \epsfxsize=6cm
\epsfbox{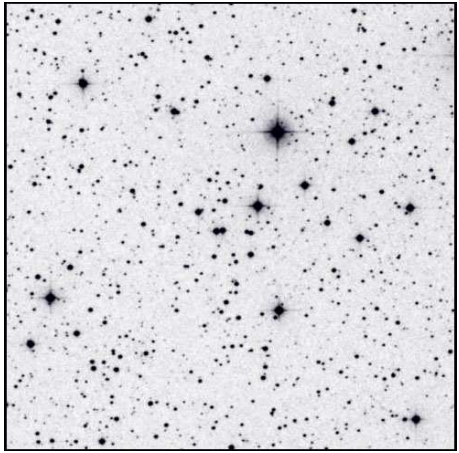}
\caption{An example for the image of NGC 7801 as taken from DSS site. North is up, east on the left.}
\label{fig-single}
\end{figure}
%---------------------------------------------------------------------
\begin{figure}
\centering \epsfxsize=8.5cm
\epsfbox{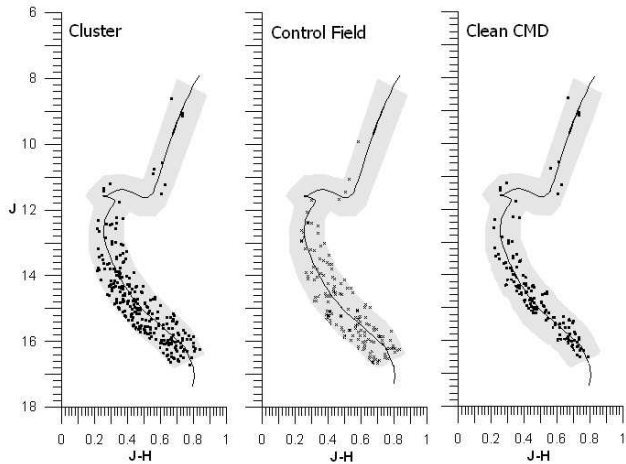}
\caption{An example for establishing a clean CMD of NGC 7801 by comparing its control field, the dark areas represent the color and magnitude filters, which applied to the cluster and its field sequences; the maximum departure is about 0.15 mag.}
\label{fig-single}
\end{figure}
%---------------------------------------------------------------------
\section{CENTER DETERMINATION}

To estimate the cluster center theoretically, it can be defined as either the center of mass or the
location of the deepest part of the gravitational potential. Observationally, the center is often
defined as the region of the highest surface brightness or the region containing the largest number
of stars (Littlefair et al. 2003). Here, the cluster center is defined as the location of maximum
density of probable member stars, after applying the color-magnitude filters. The optimized cluster
centers can be obtained by fitting a Gaussian to the profiles of star counts, in equal incremental
strips, in right ascensions and declinations. Fig. 3 represents an example for determining the
cluster center of NGC 2351. All the clusters centers are found to be in agreement with our estimations
within errors of a few arcseconds. Note that the coordinates of the centers of NGC 1857, NGC 2061,
NGC 2234, NGC 2250 and NGC 3231 are found to be different in the WEBDA and DIAS catalogs; so we used
the ``Coordinate Conversion and Precession Tool" of Chandra\footnote{\it http://cxc.harvard.edu/}
to correct their coordinates. Meanwhile, the objects NGC 2664, NGC 5385 and NGC 6863 are demonstrated
on the basis of radial velocity, original CCD photometry and proper motions to be random enhancements
of field stars (Villanova et al. 2004; Moni et al. 2010).
%----------------------------------------------------------------

\section{RADIAL DENSITY PROFILE}
The lower limit of the size of each cluster (lower border) has been obtained using the radial density profile (RDP) of the cluster's stars. The spatial coverage and the uniformity of 2MASS photometry allows one to obtain reliable data on the projected distribution of stars over extended regions around clusters (Bonatto et al. 2005).
So, within concentric shells in equal incremental steps of about 0.1$ \sim$ 0.5 arcmin from the cluster
center, the stellar density is derived out to a pre-determined radius.
Applying the empirical profile of King (1962), the cluster limiting radius can be determined at a specific
point at which it reaches a stable background density. At that radius, the JH$K_S$ photometric data would
be taken into account for estimating the main cluster properties. An example for radius determination of
NGC 1498 is shown in Fig. 4. On the other hand, the RDP of a cluster is used to evaluate the amount of
field contamination, where the clusters of field density $\leq$ 2 stars per arcmin$^{2}$ don't need to
build the RDPs or CMDs of their fields. On the contrary, for the clusters with field density $>$ 2 stars
per arcmin$^{2}$, the RDPs and CMDs of their field are needed.
%----------------------------------------------------------------
\begin{figure}
\centering \epsfxsize=10cm
\epsfbox{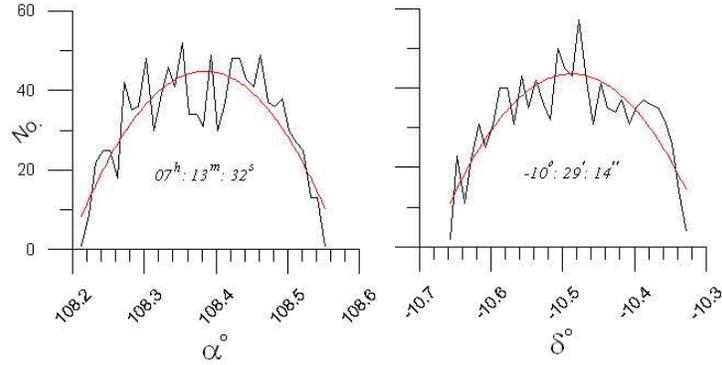}
\caption{An example for optimizing center determination for NGC 2351, the curved lines represent the Gaussian fitting profiles. Comparing the estimated coordinates with what obtained in Table 1 in the Appendix, we found that there are differences of 1 sec in $\alpha$ and 2 arcsec in $\delta$ of the cluster's center.}
\label{fig-single}
\end{figure}
%---------------------------------------------------------------------
\begin{figure}
\centering \epsfxsize=9cm
\epsfbox{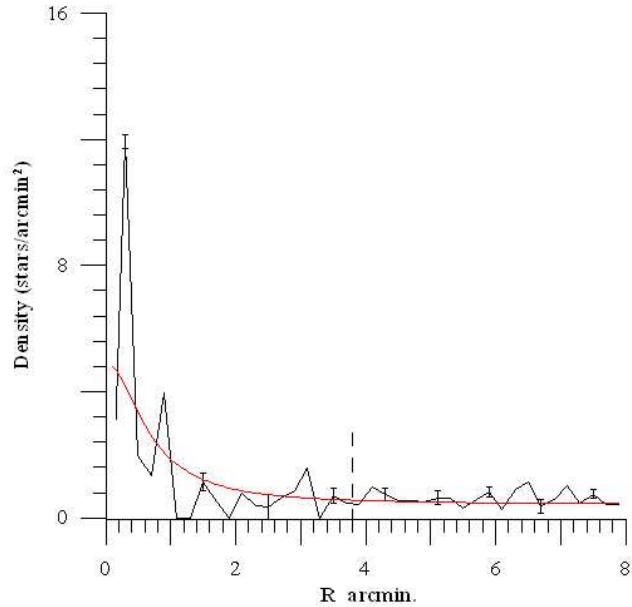}
\caption{An example for the lower radius determination for NGC 1498, the curved line represents the fitting of King (1962) model. The length of the error bars denote errors resulting sampling statistics, in accordance with Poisson distribution. The vertical dashed line refers to that point, at which the radius determine; reaching the background density of about 0.4 stars per arcmin$^{2}$.}
\label{fig-single}
\end{figure}
%---------------------------------------------------------------------
\section{FIELD-STAR DECONTAMINATION}
Usually, field stars contaminate the CMDs of a cluster, particularly at faint magnitudes and red colors. Most over-density clusters are located near the disk or/and the bulge of the Galaxy and shows crowded contaminated main sequences. These contaminated stars are always seen as a vertical redder sequence parallel to the cluster's main sequence. CMDs of such clusters surely contain field stars that might lead to artificial isochrone solutions, i.e. one can always ``fit" isochrones to such CMDs. Therefore, field-star decontamination must be used to define the intrinsic CMDs and get better isochrone fitting. To achieve this, we have to compare the CMDs of a cluster with that of a nearby control field. A control field is chosen at the same Galactic latitude, but with one degree larger or one degree smaller than that of the Galactic longitude of the cluster. From the comparison of the CMDs of such a cluster and its control field for a given magnitude and color range, we have counted the number of stars in a control field and subtracted this number from the cluster's CMDs.

It can be noted that, for a good separated cluster, the mean density of the control field is always less than the central region of the cluster. For the sample we have  investigated, we have found 12 candidates that are difficult to separate from the background field and do not have well defined RDPs or CMDs. Theses are NGC 1963, NGC 5269, NGC 6169, NGC 6415, NGC 6455, NGC 6476, NGC 6480, NGC 6529, NGC 6554, NGC 6682, NGC 6980, and NGC 6989. Most of them are located very close to the Galactic disk or/and in the direction of the Galactic bulge.
%----------------------------------------------------------------

\begin{figure}
\centering \epsfxsize=4cm
\epsfbox{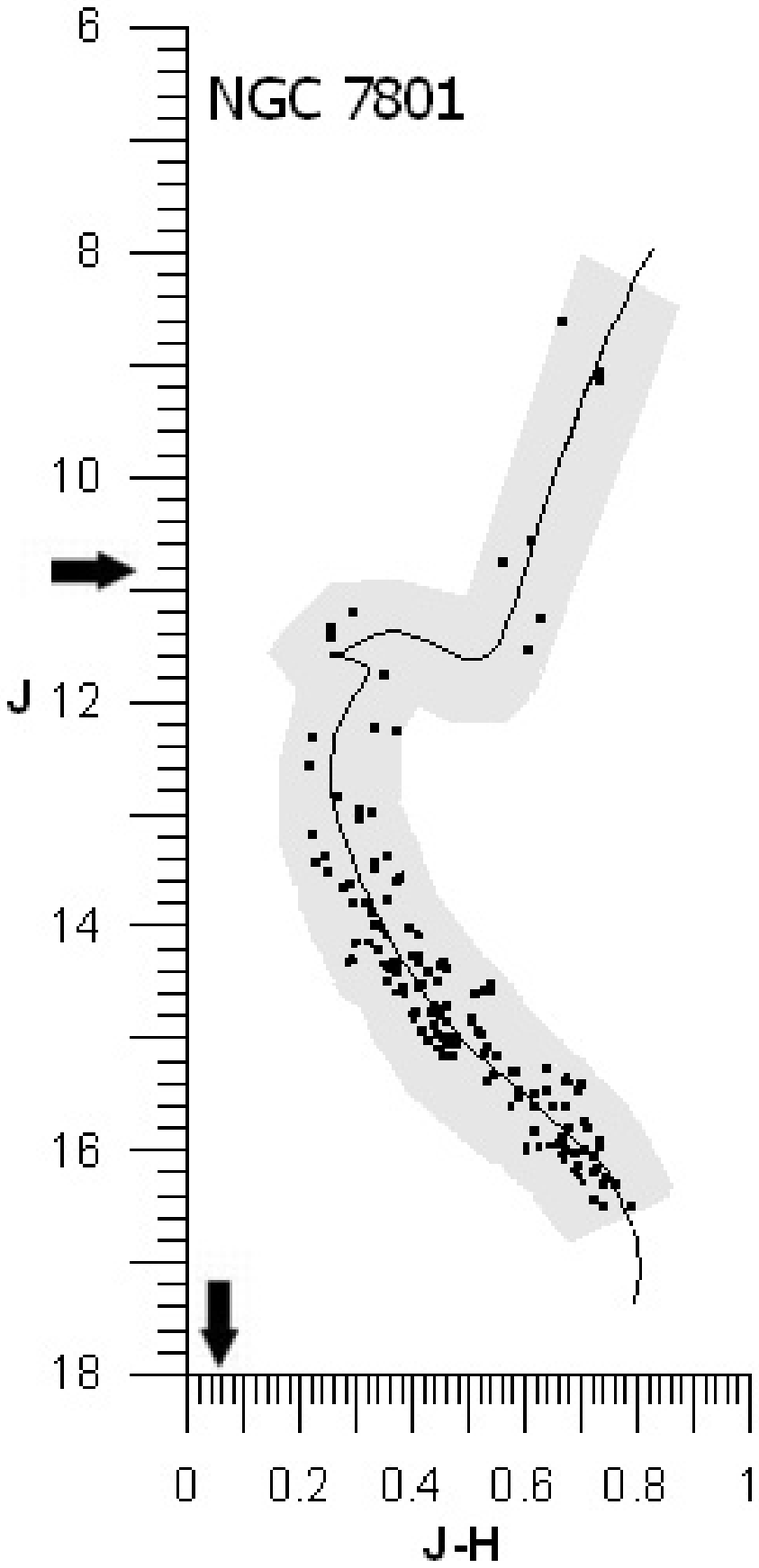}
\caption{An example for estimating the main astrophysical parameters of NGC 7801. The age, reddening and mean distance modulus are obtained from the new theoretical Padova isochrones of solar metallicity (Bonatto et al. 2004; Bica, et al. 2006) after applying the color and magnitude filters to J$\sim$(J-H) sequence, see Fig. 2. The age = 1.7 Gyr, $E_{J-H}$=0.06 mag, and $(m-M)_{J}$= 10.8 mag. The fitting errors are $\sim$ $\pm 0.10$ mag in distance modulus, and $\sim$ $\pm 0.05$ mag in color excess.}
\label{fig-single}
\end{figure}
%---------------------------------------------------------------------
\begin{figure}
\centering \epsfxsize=10cm
\epsfbox{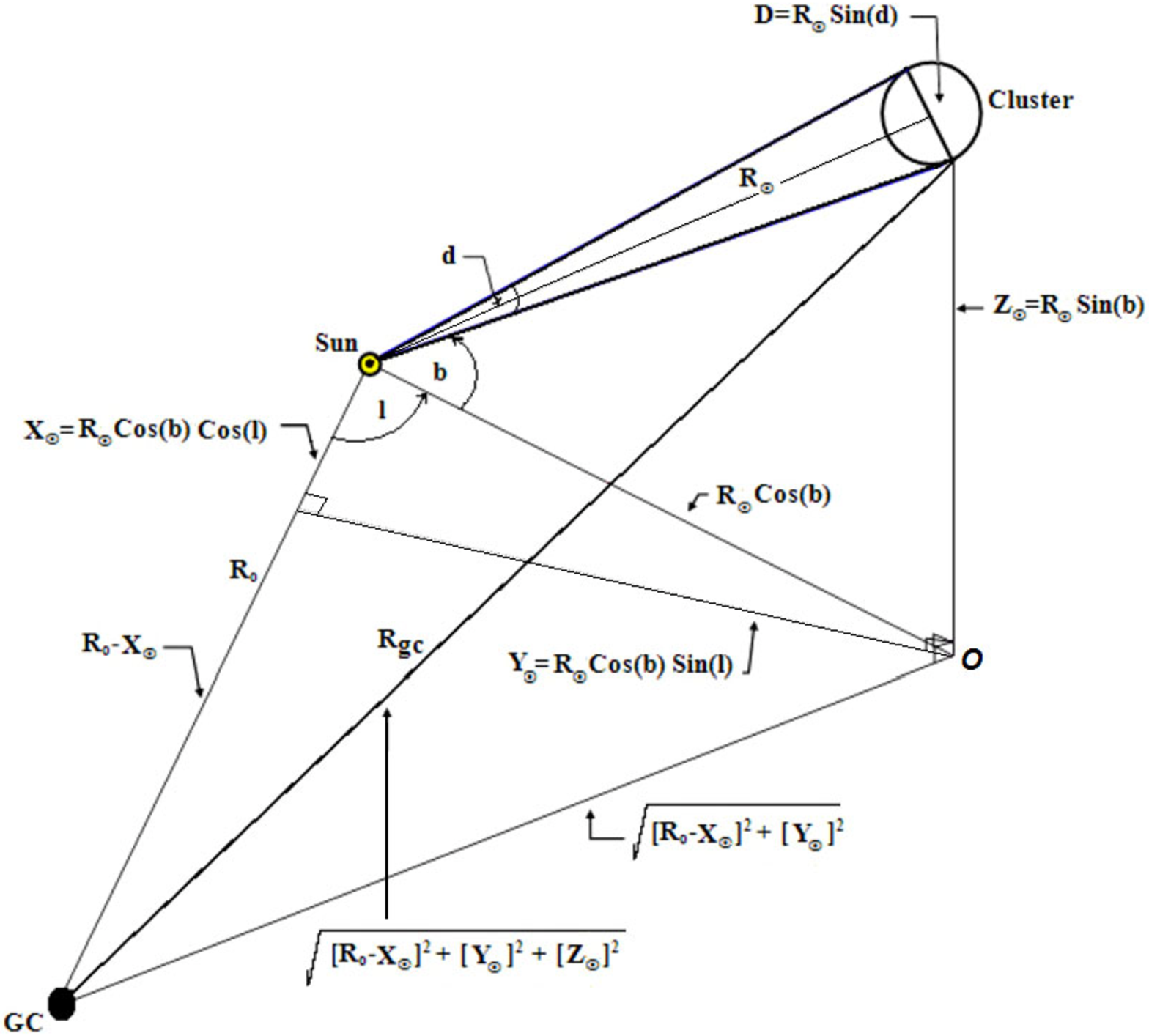}
\caption{A sketch-chart illustrates the Galactic geometric distances for a cluster in the Galaxy; taken from Tadross, A. L. (2000). Let the cluster's distances from the sun, $R_{\odot}$, from the Galactic center, $R_{gc}$, and the horizontal projected rectangular distances on the Galactic plane centered on the Sun, $X_{\odot}$, $Y_{\odot}$, and the distance from Galactic plane, $Z_{\odot}$. The sun's distance from the galactic center $R_{o}$=8.5 kpc; {\it d} \& {\it D} are the angular and linear diameter of the cluster; $\ell$ \& {\it b} are the galactic longitude and latitude of the cluster.}
\label{fig-single}
\end{figure}

\section{COLOR MAGNITUDE DIAGRAM ANALYSIS}
 CMDs are established for the stars inside radii of 1$^{'}$, 2$^{'}$, 3$^{'}$ etc. from the optimized coordinates of the centers of the clusters under investigation. We have fitted the new theoretical computed with the 2MASS $J, H$ and $K_S$ filters (Bonatto et al. 2004; Bica et al. 2006) to derive the cluster parameters. The simultaneous fittings were attempted on the $J\sim(J-H)$ and $K_S\sim(J-K_S)$ diagrams for the inner stars, which should be less contaminated by the background field.

If the number of stars are not enough for an accepted fitting, the next larger area is included, and so on. In this way, different isochrones of different ages have been applied on the CMDs of each cluster, fitting the lower envelope of the points matching the main sequence stars, turn-off point and red giant positions. Guiding by the Galactic reddening values of Schlegel et al. (1998), the average age and distance modulus are determined for each cluster. Although Schlegel's reddening values are often overestimated at low Galactic latitudes, it is still a useful source of data. Comparing our estimated reddening values with the reliable ones of Schlegel's, we found that 96\% of our sample are in agreement with Schlegel's values within ranging errors of 0.05 mag. (see Table 3 in the Appendix). The distance modulus is taken at the proper values within a ranging fitting error of about $\pm 0.10$ mag. Fig. 5 represents an example for estimating the main parameters of NGC 7801.

The observed data has been corrected for interstellar reddening using the coefficients ratios $\frac {A_{J}}{A_{V}}= 0.276$ and $\frac {A_{H}}{A_{V}}= 0.176$, which were derived from absorption rations in Schlegel et al. (1998), while the ratio $\frac {A_{K_S}}{A_{V}}= 0.118$ was derived from Dutra et al. (2002).
Therefore $\frac {E_{J-H}}{E_{B-V}}= 0.309$, $\frac {E_{J-K_S}}{E_{B-V}}= 0.488$, and then $\frac {E_{J-K_S}}{E_{J-H}}\approx$ 1.6 $\pm$ 0.15 can be derived easily from the above ratios, where R$_{V}=\frac {A_{V}}{E_{B-V}}= 3.1$.
%---------------------------------------------------------------------

\section{GALACTIC GEOMETRIC DISTANCES}

Galactic geometric distances are the distance from the sun, $R_{\odot}$, distance from the Galactic center, $R_{gc}$, and the projected rectangular distances on the Galactic plane centered on the Sun, $X_{\odot}$, $Y_{\odot}$, as well as the distance from Galactic plane, $Z_{\odot}$. The importance of such distances is that they can be used to investigate the geometry of the Galaxy or/and the traces of the Milky-Way arms. Fig. 6 represents a sketch-chart for the calculation of the geometric distances for a cluster in the Galaxy, taken from Tadross (2000).
%---------------------------------------------------------------------
\begin{figure*}
\centering \epsfxsize=10cm
\epsfbox{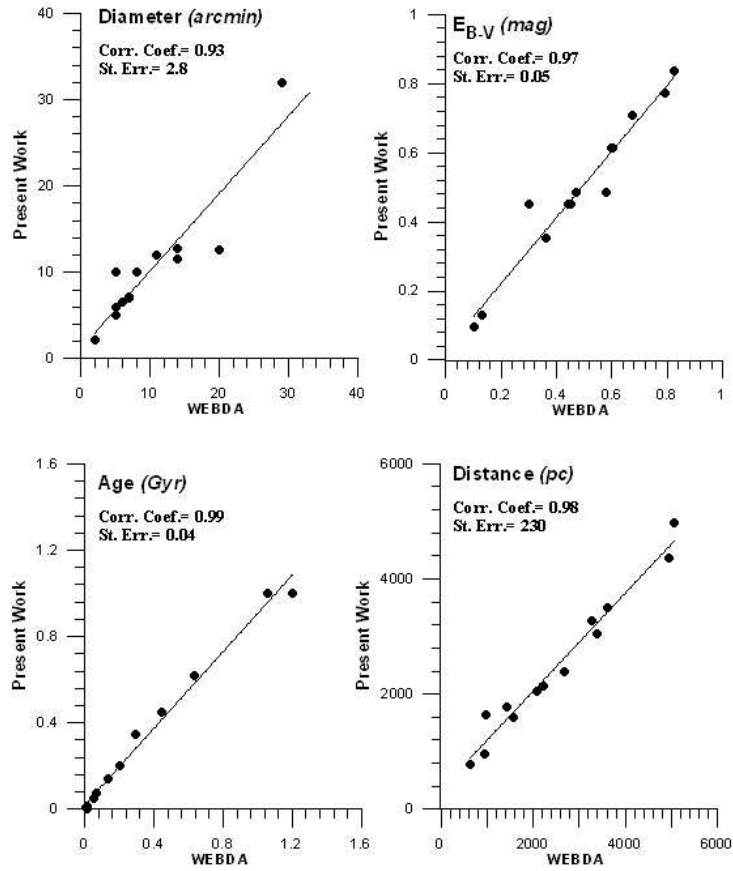}
\caption{The comparison results of the main parameters of thirteen clusters with those in the literature. Diameters, color excesses, ages and distances with the correlation coefficient and standard error for each relation are obtained.}
\label{fig-single}
\end{figure*}

\section{CONCLUSIONS}

Following the above procedures, the basic astrophysical parameters, namely, angular diameter, age, reddening, distance modulus and the Galactic geometric distances, have been obtained for 120 NGC previously unstudied open clusters. The results may be used to investigate, for instance, the cluster formation rate, the geometry of the Galaxy, the time scale for cluster dissolution, etc. The same procedures have been applied to re-estimate 13 previously studied NGC clusters for calibration purposes. It must be mentioned that some distant clusters which lie behind or embedded within dark clouds of gas and dust are influenced by differential reddening that lead to some biases in estimating parameters from optical data. Such clusters are heavily obscured in the optical, but easily detected in the infrared band. On the other hand, it is difficult to detect the cluster members in the infrared band, especially those with low stellar density, because of the huge background field. Therefore, the difficulties in the detection and extraction of cluster members increasingly worsen towards NIR bands (Zakharova \& Loktin 2006).

A Comparison of the astrophysical parameters of the thirteen clusters that overlap between the present work and those in WEBDA are listed in Table 2 in the Appendix. The relations between the present results and those obtained from WEBDA for diameter, color excess, age and distance can be seen in Fig. 7. The correlation coefficient and standard error for each relation are shown in the same figure. We can see that the derived parameters are reasonable and very close to the published ones, which indicates that our reduction procedure is very reliable. The final results of the investigated 133 star clusters that investigated here are listed in Table 3 in the Appendix. All the CMDs of our sample are electronically available on demand.\\
%The CMDs and isochrone fits for all clusters are electronically available when demanded.
%---------------------------------------------------------------------
\section*{Acknowledgments}
This publication makes use of data products from the Two Micron All Sky Survey {\it 2MASS}, which is a joint project of the University of Massachusetts and the Infrared Processing and Analysis Center/California Institute of Technology, funded by the National Aeronautics and Space Administration and the National Science Foundation. Catalogues from {\it CDS}/{\it SIMBAD} (Strasbourg), and Digitized Sky Survey {\it DSS} images from the Space Telescope Science Institute have been employed.

\section*{References}

Bica, E., Bonatto, Ch., Blumberg, R. 2006, A\&A, 460, 83\\
Bonatto, Ch., Bica, E., Girardi, L. 2004, A\&A, 415, 571\\
Bonatto, Ch., Bica, E., Santos, J. 2005, A\&A, 433, 917\\
Chen, L., Hou, J., Wang, J. 2003, AJ, 125, 1397\\
Dutra, C., Santiago, B., Bica, E. 2002, A\&A, 381, 219\\
Glushkova, E., Koposov, S., Zolotukhin, I. 2008, A\&A, 486, 771\\
Hou, J., Prantzos, N., Boissier, S. 2000, A\&A, 362, 921\\
Kim, S., Sung, H. 2003, JKAS, 36, 13\\
Kim, S., et al. 2005, JKAS, 38, 429\\
King, I. 1962, AJ, 67, 471\\
Littlfair, S. et al. 2003, MNRAS. 345, 1205\\
Moni, C. et al. 2010, A\&A, 510, 44\\
Piskunov, A., et al. 2006, A\&A, 445, 545\\
Schlegel, D., Finkbeiner, D., Davis, M. 1998, ApJ, 500, 525\\
Skrutskie, M., Cutri, R., Stiening, R., et al. 2006, AJ, 131, 1163\\
Tadross, A. L. 2000,~ PhD thesis, Faculty of Science, Cairo University, pp. 15\\
Tadross, A. L. 2003, NewA, 8, 737\\
Tadross, A. L. 2008{\it a}, NewA, 13, 370\\
Tadross, A. L. 2008{\it b}, MNRAS, 389, 285\\
Tadross, A. L. 2009{\it a}, NewA, 14, 200\\
Tadross, A. L. 2009{\it b}, Ap\&SS, 323, 383\\
Villanova, S. et al. 2004, A\&A, 428, 67\\
Zakharova, P. E., Loktin, A. V. 2006, A\&AT, 25, 171\\

\begin{table}
\begin{center}
\centering
\caption{The equatorial, Galactic positions and the diameters of the investigated clusters, as taken from ``Webda" and ``Dias"; sorted by right ascensions. \label{tbl1}}
\bigskip
\scriptsize
\doublerulesep2.0pt
\renewcommand\arraystretch{1.3}
\begin{tabular}{ccccccc}
\hline \hline
Index
            & Cluster
            & $\alpha~^{h}~^{m}~^{s}$
            & $\delta~^{\circ}~{'}~{''}$
            & G. Long.$^{\circ}$
            & G. Lat.$^{\circ}$
            & Diam.$^{'}$\\
            \hline
1   & NGC 7801 &  00:00:21  &  +50:44:30 & 114.73 & -11.315 & 8 \\
2   & NGC 7826 &  00:05:17  &  -20:41:30 & 61.875 & -77.653 & 20 \\
3   & NGC 7833 &  00:06:31  &  +27:38:30 & 110.9 & -34.178  & 1.3 \\
4   & NGC ~110 &  00:27:25  &  +71:23:00 & 121 & 8.602 & 20 \\
5   & NGC ~272 &  00:51:24  &  +35:49:54 & 122.93 & -27.05 & 4 \\
6   & NGC ~305 &  00:56:20  &  +12:04:00 & 124.83 & -50.787 & 6 \\
7   & NGC ~657 &  01:43:21  &  +55:50:11 & 130.22 & -6.299 & 4 \\
8   & NGC ~743 &  01:58:37  &  +60:09:18 & 131.2 & -1.633 & 6 \\
9   & NGC ~956 &  02:32:30  &  +44:35:36 & 141.18 & -14.624 & 9 \\
10  & NGC 1520 &  03:57:51  &  -76:47:42 & 291.14 & -35.705 & 5 \\
11  & NGC 1498 &  04:00:18  &  -12:00:54 & 203.62 & -43.33 & 2 \\
12  & NGC 1557 &  04:13:11  &  -70:28:18 & 283.77 & -38.261 & 21 \\
13  & NGC 1785 &  04:58:35  &  -68:50:40 & 280.01 & -35.247 & 3 \\
14  & NGC 1724 &  05:03:32  &  +49:29:30 & 158.45 & 4.845 & 2 \\
15  & NGC 1807 &  05:10:43  &  +16:31:18 & 186.09 & -13.495 & 15 \\
16  & ~NGC 1857$^{*}$ &  05:20:12  &  +39:21:00 & 168.41 & 1.279 & 6 \\
17  & NGC 1891 &  05:21:25  &  -35:44:24 & 239.7 & -32.878 & 15 \\
18  & ~NGC 1963$^{x}$ &  05:32:17  &  -36:23:30 & 240.99 & -30.869 & 13 \\
19  & NGC 2017 &  05:39:17  &  -17:50:48 & 221.62 & -23.708 & 6 \\
20  & ~NGC 2061$^{*}$ &  05:42:42  &  -34:00:34 & 238.92 & -28.231 & 16 \\
21  & NGC 2026 &  05:43:12  &  +20:08:00 & 187.23 & -5.059 & 11 \\
22  & NGC 2039 &  05:44:00  &  +08:41:30 & 197.58 & -10.274 & 30 \\
23  & NGC 2013 &  05:44:01  &  +55:47:36 & 156.51 & 13.417 & 4 \\
24  & NGC 2063 &  05:46:43  &  +08:46:54 & 197.29 & -10.766 & 9 \\
25  & NGC 2132 &  05:55:18  &  -59:54:36 & 268.69 & -30.2 & 17 \\
26  & NGC 2165 &  06:11:04  &  +51:40:36 & 162.16 & 15.13 & 9 \\
27  & NGC 2189 &  06:12:09  &  +01:03:54 & 207.46 & -8.24 & 7 \\
28  & NGC 2220 &  06:21:11  &  -44:45:30 & 252.5 & -23.926 & 13 \\
29  & NGC 2219 &  06:23:44  &  -04:40:36 & 213.96 & -8.288 & 6 \\
30  & NGC 2224 &  06:27:28  &  +12:35:36 & 198.97 & 0.544 & 13 \\
31  & ~NGC 2234$^{*}$ &  06:29:24  &  +16:41:00 & 195.61 & 2.812 & 25 \\
32  & ~NGC 2250$^{*}$ &  06:32:48  &  -05:02:00 & 215.31 & -6.432 & 3 \\
33  & NGC 2248 &  06:34:35  &  +26:18:16 & 187.54 & 8.247 & 1.5 \\
34  & NGC 2260 &  06:38:03  &  -01:28:24 & 212.72 & -3.65 & 18 \\
35  & NGC 2265 &  06:41:41  &  +11:54:18 & 201.23 & 3.268 & 9 \\
36  & NGC 2312 &  06:58:47  &  +10:17:42 & 204.56 & 6.29 & 6 \\
37  & NGC 2318 &  06:59:27  &  -13:41:54 & 226.05 & -4.46 & 12 \\
38  & NGC 2348 &  07:03:03  &  -67:24:42 & 278.14 & -23.809 & 18 \\
39  & NGC 2331 &  07:06:59  &  +27:15:42 & 189.73 & 15.218 & 14 \\
40  & NGC 2338 &  07:07:47  &  -05:43:12 & 219.89 & 1.012 & 3 \\
41  & NGC 2349 &  07:10:48  &  -08:35:36 & 222.78 & 0.351 & 10 \\
42  & NGC 2352 &  07:13:05  &  -24:02:18 & 236.77 & -6.263 & 6 \\
43  & NGC 2351 &  07:13:31  &  -10:29:12 & 224.77 & 0.068 & 4 \\
44  & NGC 2364 &  07:20:46  &  -07:33:00 & 223.0 & 3.02 & 6 \\
45  & NGC 2408 &  07:40:09  &  +71:39:18 & 143.63 & 29.466 & 24 \\
\hline
\end{tabular}
\end{center}
\end{table}

\begin{table}
\begin{center}
\scriptsize
%\centerline{
\centering
\renewcommand\arraystretch{1.3}
\begin{tabular}{ccccccc}
& & & {\bf Table 1.} --- {Continued} & & & \\ \\
\hline \hline
Index
            & Cluster
            & $\alpha~^{h}~^{m}~^{s}$
            & $\delta~^{\circ}~{'}~{''}$
            & G. Long.$^{\circ}$
            & G. Lat.$^{\circ}$
            & Diam.$^{'}$\\
            \hline
46  & NGC 2455 &  07:49:01  &  -21:18:06 & 238.35 & 2.32 & 5 \\
47  & NGC 2459 &  07:52:02  &  +09:33:24 & 211.11 & 17.762 & 1 \\            
48  & NGC 2587 &  08:23:25  &  -29:30:30 & 249.46 & 4.472 & 2 \\
49  & NGC 2609 &  08:29:32  &  -61:06:36 & 276.15 & -12.754 & 5 \\
50  & NGC 2666 &  08:49:47  &  +44:42:12 & 175.92 & 39.278 & 11 \\
51  & NGC 2678 &  08:50:02  &  +11:20:18 & 216.04 & 31.421 & 10 \\
52  & NGC 2932 &  09:35:28  &  -46:48:36 & 271.76 & 3.855 & 20 \\
53  & NGC 2995 &  09:44:04  &  -54:46:48 & 278.05 & -1.228 & 4.5 \\
54  & ~NGC 3231$^{*}$ &  10:26:58  &  +66:48:55 & 141.96 & 44.604 & 2.5 \\
55  & NGC 3446 &  10:52:12  &  -45:08:54 & 281.86 & 12.79 & 15 \\
56  & NGC 3520 &  11:07:08  &  -18:01:24 & 270.77 & 38.25 & 0.6 \\
57  & NGC 3909 &  11:49:49  &  -48:15:06 & 292.46 & 13.363 & 15 \\
58  & NGC 4230 &  12:17:20  &  -55:06:06 & 298.03 & 7.445 & 8 \\
59  & NGC 5155 &  13:29:35  &  -63:25:30 & 307.18 & -0.869 & 12 \\
60  & NGC 5269 &  13:44:44  &  -62:54:54 & 308.96 & -0.668 & 3 \\
61  & ~NGC 5284$^{x}$ &  13:46:41  &  -59:12:00 & 309.96 & 2.916 & 30 \\
62  & NGC 5299 &  13:50:26  &  -59:56:54 & 310.26 & 2.081 & 25 \\
63  & NGC 5381 &  14:00:41  &  -59:35:12 & 311.6 & 2.114 & 11 \\
64  & NGC 5800 &  15:01:47  &  -51:55:06 & 322.44 & 5.946 & 12 \\
65  & NGC 5925 &  15:27:26  &  -54:31:42 & 324.36 & 1.72 & 20 \\
66  & NGC 5998 &  15:49:34  &  -28:35:18 & 343.82 & 19.809 & 4 \\
67  & ~NGC 6169$^{x}$ &  16:34:04  &  -44:02:42 & 339.38 & 2.515 & 12 \\
68  & NGC 6334 &  17:20:49  &  -36:06:12 & 351.15 & 0.475 & 20 \\
69  & NGC 6360 &  17:24:27  &  -29:52:18 & 356.73 & 3.13 & 5 \\
70  & NGC 6357 &  17:24:43  &  -34:12:06 & 353.17 & 0.895 & 2 \\
71  & NGC 6374 &  17:32:18  &  -32:36:00 & 355.38 & 0.465 & 4 \\
72  & ~NGC 6415$^{x}$ &  17:44:18  &  -35:04:00 & 354.41 & -2.753 & ? \\
73  & NGC 6421 &  17:45:44  &  -33:41:36 & 355.9  & -2.405 & 8 \\
74  & NGC 6437 &  17:48:24  &  -35:21:00 & 354.44 & -4.184 & 15 \\
75  & ~NGC 6455$^{x}$ &  17:51:08  &  -35:20:18 & 355.32 & -4.33 & 8 \\
76  & ~NGC 6476$^{x}$ &  17:54:02  &  -29:08:42 & 0.805 & -1.704 & ? \\
77  & ~NGC 6480$^{x}$ &  17:54:26  &  -30:27:06 & 359.7 & -2.44 & 5 \\
78  & NGC 6525 &  18:02:06  &  +11:01:24 & 37.378 & 15.89 & 8 \\
79  & ~NGC 6529$^{x}$ &  18:05:29  &  -36:17:48 & 355.71 & -7.3 & 16 \\
80  & ~NGC 6554$^{x}$ &  18:08:59  &  -18:26:06 & 11.762 & 0.648 & 20 \\
81  & NGC 6573 &  18:13:41  &  -22:07:06 & 9.063 & -2.091 & 1 \\
82  & NGC 6595 &  18:17:04  &  -19:51:54 & 11.422 & -1.713 & 4 \\
83  & NGC 6605 &  18:18:21  &  -14:56:42 & 15.9 & 0.35 & 15 \\
84  & NGC 6588 &  18:20:33  &  -63:48:30 & 330.84 & -20.878 & 5 \\
85  & NGC 6659 &  18:33:59  &  +23:35:42 & 52.48 & 14.156 & 7 \\
86  & ~NGC 6682$^{x}$ &  18:39:37  &  -04:48:48 & 27.302 & 0.427 & 19 \\
87  & NGC 6698 &  18:48:04  &  -25:52:42 & 9.255 & -10.792 & 10 \\
88  & NGC 6724 &  18:56:46  &  +10:25:42 & 42.842 & 3.577 & 3 \\
89  & NGC 6735 &  19:00:37  &  -00:28:30 & 34.402 & -1.825 & 12 \\
90  & NGC 6743 &  19:01:20  &  +29:16:36 & 60.367 & 10.922 & 7 \\
91  & NGC 6773 &  19:15:03  &  +04:52:54 & 39.976 & -3.003 & 8 \\
92  & NGC 6775 &  19:16:48  &  -00:55:24 & 36.657 & -5.217 & 2 \\
\hline
\end{tabular}
\end{center}
\end{table}

\begin{table}
\begin{center}
\scriptsize
%\centerline{
\centering
\renewcommand\arraystretch{1.3}
\begin{tabular}{ccccccc}
& & & {\bf Table 1.} --- {Continued} & & & \\ \\
\hline \hline
Index
            & Cluster
            & $\alpha~^{h}~^{m}~^{s}$
            & $\delta~^{\circ}~{'}~{''}$
            & G. Long.$^{\circ}$
            & G. Lat.$^{\circ}$
            & Diam.$^{'}$\\
            \hline
93  & NGC 6795 &  19:26:22  &  +03:30:54 & 40.077 & -6.137 & 8 \\
94  & NGC 6815 &  19:40:44  &  +26:45:30 & 62.135 & 2.045 & 30 \\
95  & NGC 6832 &  19:48:15  &  +59:25:18 & 92.005 & 16.349 & 15 \\
96  & NGC 6837 &  19:53:08  &  +11:41:54 & 50.519 & -8.009 & 3 \\
97  & NGC 6839 &  19:54:33  &  +17:56:18 & 56.114 & -5.152 & 6 \\
98  & NGC 6840 &  19:55:18  &  +12:07:36 & 51.162 & -8.253 & 6 \\
99  & NGC 6843 &  19:56:06  &  +12:09:48 & 51.293 & -8.404 & 5 \\
100 & NGC 6846 &  19:56:28  &  +32:20:54 & 68.691 & 1.919 & 2 \\
101 & NGC 6847 &  19:56:37  &  +30:12:48 & 66.882 & 0.783 & 30 \\
102 & NGC 6856 &  19:59:17  &  +56:07:48 & 89.683 & 13.527 & 2 \\
103 & NGC 6858 &  20:02:56  &  +11:15:30 & 51.36  & -10.305 & 10 \\
104 & NGC 6859 &  20:03:49  &  +00:26:36 & 41.812 & -15.836 & 8  \\
105 & NGC 6873 &  20:07:13  &  +21:06:06 & 60.451 & -6.154  & 15 \\
106 & NGC 6895 &  20:16:29  &  +50:13:48 & 85.887 &  8.281  & 30 \\
107 & NGC 6904 &  20:21:48  &  +25:44:24 & 66.135 & -6.311  & 8 \\
108 & NGC 6938 &  20:34:42  &  +22:12:54 & 64.91  & -10.743 & 7 \\
109 & NGC 6950 &  20:41:04  &  +16:37:06 & 61.107 & -15.198 & 15 \\
110 & ~NGC 6980$^{x}$ &  20:52:48  &  -05:50:12 & 42.105 & -29.578 & 10 \\
111 & ~NGC 6989$^{x}$ &  20:54:06  &  +45:14:24 & 85.654 &  0.238  & 10 \\
112 & NGC 7023 &  21:01:35  &  +68:10:12 & 104.06 & 14.19   & 5 \\
113 & NGC 7011 &  21:01:49  &  +47:21:12 & 88.126 & 0.607   & 3 \\
114 & NGC 7005 &  21:01:57  &  -12:52:50 & 35.81  & -34.711 & 1.5 \\
115 & NGC 7024 &  21:06:09  &  +41:29:18 & 84.265 & -3.877  & 5 \\
116 & NGC 7037 &  21:10:54  &  +33:45:48 & 79.133 & -9.761  & 6 \\
117 & NGC 7050 &  21:15:12  &  +36:10:24 & 81.533 & -8.774  & 7 \\
118 & NGC 7055 &  21:19:30  &  +57:34:12 & 97.449 & 5.597   & 3 \\
119 & NGC 7071 &  21:26:39  &  +47:55:12 & 91.426 & -2.024  & 3 \\
120 & NGC 7084 &  21:32:33  &  +17:30:30 & 69.963 & -24.302 & 16 \\
121 & NGC 7093 &  21:34:21  &  +45:57:54 & 91.043 & -4.348  & 9 \\
122 & NGC 7129 &  21:42:59  &  +66:06:48 & 105.4  & 9.885   & 7 \\
123 & NGC 7127 &  21:43:41  &  +54:37:48 & 97.907 & 1.15    & 5 \\
124 & NGC 7134 &  21:48:55  &  -12:58:24 & 41.98  & -45.141 & 1 \\
125 & NGC 7175 &  21:58:46  &  +54:49:06 & 99.717 & -0.075  & 29 \\
126 & NGC 7193 &  22:03:03  &  +10:48:06 & 70.094 & -34.279 & 13 \\
127 & NGC 7352 &  22:39:43  &  +57:23:42 & 105.9  & -1.054  & 5 \\
128 & NGC 7394 &  22:50:23  &  +52:08:06 & 104.79 & -6.422  & 9 \\
129 & NGC 7429 &  22:56:00  &  +59:58:24 & 108.96 &  0.272  & 14 \\
130 & NGC 7686 &  23:30:07  &  +49:08:00 & 109.51 & -11.608 & 14 \\
131 & NGC 7708 &  23:35:01  &  +72:50:00 & 117.4  & 10.788  & 23 \\
132 & NGC 7795 &  23:58:37  &  +60:02:06 & 116.38 & -2.163  & 21 \\ \hline
\end{tabular}
\end{center}
\end{table}

%--------------------------------------------------------------------------------------
\begin{table}
\begin{center}
\centering
\caption{The available parameters of the calibrated clusters as obtained from ``Webda". \label{tbl1}}
\scriptsize
\doublerulesep2.0pt
\renewcommand\arraystretch{1.3}
\begin{tabular}{cccccccccc}
\hline \hline
Index & Cluster  & $\alpha~^{h}~^{m}~^{s}$  & $\delta~^{\circ}~{'}~{''}$
& G. Long.$^{\circ}$  & G. Lat.$^{\circ}$ & Dist. & E(B-V) & Age & Diam.$^{'}$\\
            &&&&&& \it pc.& \it mag.& \it G yr.& \it arcmin.\\ \hline
1   & NGC ~133 &  00:31:19  &  +63:21:00  & 120.678 & ~0.566 & ~630 & 0.60 & 0.01 & 7 \\
2   & NGC 1893 &  05:22:44  &  +33:24:42  & 173.585 & --1.68 & 3280 & 0.58 & 0.01 & 11 \\
3   & NGC 2158 &  06:07:25  &  +24:05:48  & 186.634 & 1.781  & 5071 & 0.36 & 1.05 & 5 \\
4   & NGC 2266 &  06:43:19  &  +26:58:12  & ~187.79 & 10.294  & 3400 & 0.10 & 0.63 & 5 \\
5   & NGC 2588 &  08:23:10  &  --32:58:30 & ~252.28 & ~2.449  & 4950 & 0.30 & 0.45 & 2 \\
6   & NGC 3496 &  10:59:36  &  --60:20:12 & 289.515 & --0.411 & ~990 & 0.47 & 0.30 & 8 \\
7   & NGC 6005 &  15:55:48  &  --57:26:12 & ~325.78 & --2.986 & 2690 & 0.45 & 1.20 & 5 \\
8   & NGC 6451 &  17:50:41  &  --30:12:36 & 359.478 & --1.601 & 2080 & 0.67 & 0.14 & 7 \\
9   & NGC 6603 &  18:18:26  &  --18:24:24 & 12.86   & --1.306 & 3600 & 0.79 & 0.20 & 6 \\
10  & NGC 6755 &  19:07:49  &  +04:16:00  & ~38.598 & --1.688 & 1421 & 0.83 & 0.05 & 14 \\
11  & NGC 6871 &  20:05:59  &  +35:46:36  & ~72.645 & ~~2.054 & 1574 & 0.44 & 0.01 & 29 \\
12  & NGC 7039 &  21:10:48  &  +45:37:00  & 87.879  & --1.705 & ~951 & 0.13 & 0.07 & 14 \\
13  & NGC 7380 &  22:47:21  &  +58:07:54  & 107.141 & --0.884 & 2222 & 0.60 & 0.01 & 20 \\
\hline
\end{tabular}
\end{center}
%\begin{tabnote}
%\hskip18pt $^{\rm a}$ Sample table footnote. \\
%\end{tabnote}
%\begin{tabnote}
%\hskip18pt $^{\rm b}$ Sample table footnote. \\
%\end{tabnote}
\end{table}
%--------------------------------------------------------------------------------------
\begin{table}
\begin{center}
\caption{The derived astrophysical parameters for the investigated clusters. Columns display, respectively, index, cluster name, angular diameter, age, reddening, Schlegel's et al. reddening values, distance modulus, distance from the sun, distance from the Galactic center, the projected distances on the Galactic plane from the sun, and the distance from Galactic plane.}
\bigskip
\scriptsize
%\centerline{
\centering
\setlength{\tabcolsep}{0.6mm}
\renewcommand\arraystretch{1.2}
\begin{tabular}{ccccccccccccc}
\hline \hline
Index   & Cluster  & Diam.  & Age  & E$_{B-V}$  & Sch.  & m-M  & Dist.  & R$_{gc}$ & X$_{\odot}$  & Y$_{\odot}$  & Z$_{\odot}$\\
            &   & \it arcmin & \it Gyr & \it mag  & \it mag  & \it mag  & \it pc  & \it kpc   & \it pc & \it pc  & \it pc\\
\hline
1  & NGC 7801 & ~8.0 & 1.7 $^{\pm0.12}$ & 0.17 $^{\pm0.05}$ & 0.19  & 10.7 $^{\pm0.1~\Downarrow}$ & 1275 $^{\pm60}$ & 9.11  & 523   & 1136   & --250 \\
2  & NGC 7826 & 20.0 & 2.2 $^{\pm0.09}$ & 0.03 $^{\pm0.01}$ & 0.02  & 9.00 & ~620 $^{\pm29}$ & 8.23  & --62  & 117    & --606 \\
3  & NGC 7833 & ~2.6 & 2.0 $^{\pm0.08}$ & 0.06 $^{\pm0.02}$ & 0.07  & 10.8 & 1410 $^{\pm65}$ & 9.10  & 416   & 1090   & --792 \\
4  & NGC ~110 & 22.0 & 0.9 $^{\pm0.04}$ & 0.46 $^{\pm0.10}$ & 0.44  & 10.7 & 1150 $^{\pm53}$ & 9.14  & 585   & 975   & 172 \\
5  & ~NGC ~133$^{c}$ &  ~7.0 &  0.01 $^{\pm0.00}$ & 0.61 $^{\pm0.10}$ & 1.45 $\dag$ & 10.0 & ~780 $^{\pm36}$ & ~8.92 & ~398 & ~671 & ~~8 \\
6   & NGC ~272  & ~6.4  & 2.5 $^{\pm0.10}$ & 0.06 $^{\pm0.02}$ & 0.05 & 10.2 & 1068 $^{\pm50}$ & 9.12  & 517   & 798   & --486 \\
7   & NGC ~305  & ~7.0  & 2.0 $^{\pm0.08}$ & 0.06 $^{\pm0.02}$ & 0.07 & 10.9 & 1475 $^{\pm68}$ & 9.42  & 533   & 765   & --1143 \\
8   & NGC ~657  & ~6.0  & 1.6 $^{\pm0.11}$ & 0.34 $^{\pm0.05}$ & 0.36 & 11.0 & 1372 $^{\pm63}$ & 9.44  & 881   & 1041  & --151 \\
9   & NGC ~743  & ~8.0  & 0.5 $^{\pm0.02}$ & 0.95 $^{\pm0.20}$ & 0.96 & 11.9 & 1618 $^{\pm75}$ & 9.64  & 1065  & 1217  & --46 \\
10  & NGC ~956  & 10.0 & 1.0 $^{\pm0.04}$ & 0.10 $^{\pm0.05}$ & 0.09 & 10.9 & 1455 $^{\pm67}$ & 9.68  & 1097  & 883    & --367 \\
11  & NGC 1520 & ~7.2  & 2.0 $^{\pm0.08}$ & 0.06 $^{\pm0.01}$ & 0.08 & 9.50 & ~775 $^{\pm36}$ & 8.25  & --227 & --587  & --452 \\
12  & NGC 1498 & ~7.6  & 1.6 $^{\pm0.11}$ & 0.04 $^{\pm0.02}$ & 0.05 & 10.1 & 1020 $^{\pm47}$ & 9.44  & 680  & --297  & --700 \\
13  & NGC 1557 & 26.0 & 3.0 $^{\pm0.12}$ & 0.11 $^{\pm0.05}$ & 0.10 & 10.2 & 1055 $^{\pm49}$ & 8.31  & --197 & --805  & --653 \\
14  & NGC 1785 & ~6.0  & 0.5 $^{\pm0.02}$ & 0.06 $^{\pm0.02}$ & 0.08 & 12.5 & 3080 $^{\pm140}$ & 8.52  & --437 & --2477 & --1777 \\
15  & NGC 1724 & 16.0 & 0.6 $^{\pm0.02}$ & 0.57 $^{\pm0.10}$ & 0.58 & 11.3 & 1437 $^{\pm66}$ & 9.85  & 1332  & 526    & 121 \\
16  & NGC 1807 & 18.0 & 1.0 $^{\pm0.04}$ & 0.32 $^{\pm0.05}$ & 0.30 & 10.2 & ~960 $^{\pm44}$ & 9.46  & 928   & --99   & --224 \\
17  & NGC 1857 & ~8.0  & 0.16 $^{\pm0.12}$ & 0.97 $^{\pm0.20}$ & 0.99 & 11.8 & 1545 $^{\pm71}$ & 10.02 & 1513  & 310    & 34 \\
18  & NGC 1891 & 20.0 & 2.0 $^{\pm0.05}$ & 0.03 $^{\pm0.02}$ & 0.04 & 9.70 & ~860 $^{\pm40}$ & 8.96  & 364   & --624  & --467 \\
19  & ~NGC 1893$^{c}$ &  12.0 & 0.005 $^{\pm0.00}$ & 0.48 $^{\pm0.05}$ & 2.13 $\dag$ & 13.0 & 3270 $^{\pm150}$ & 11.75 & ~3248& ~365 & ~--95 \\
20  & NGC 2017 & 12.0 & 1.6 $^{\pm0.11}$ & 0.06 $^{\pm0.02}$ & 0.06 & 10.3 & 1120 $^{\pm52}$ & 9.37  & 767   & --681  & --450 \\
21  & NGC 2061 & 18.0 & 2.1 $^{\pm0.08}$ & 0.03 $^{\pm0.01}$ & 0.04 & 8.70 & ~542 $^{\pm25}$ & 8.79  & 246   & --409  & --256 \\
22  & NGC 2026 & 16.0 & 0.55 $^{\pm0.02}$ & 0.60 $^{\pm0.10}$ & 0.59 & 11.3 & 1418 $^{\pm65}$ & 9.91  & 1401  & --178  & --125 \\
23   & NGC 2039 & 10.0 & 1.2 $^{\pm0.05}$ & 0.32 $^{\pm0.05}$ & 0.33 & 10.1 & ~920 $^{\pm42}$ & 9.38  & 863   & --273  & --164 \\
24   & NGC 2013 & ~6.0  & 1.5 $^{\pm0.06}$ & 0.23 $^{\pm0.05}$ & 0.21 & 10.4 & 1100 $^{\pm51}$ & 9.52  & 981   & 426    & 255 \\
25   & NGC 2063 & 12.0 & 1.3 $^{\pm0.05}$ & 0.32 $^{\pm0.05}$ & 0.33 & 11.2 & 1525 $^{\pm70}$ & 9.97  & 1430  & --445  & --285 \\
26   & NGC 2132 & 24.0 & 1.65 $^{\pm0.12}$ & 0.06 $^{\pm0.02}$ & 0.05 & 10.0 & ~974 $^{\pm45}$ & 8.58  & 19   & --842  & --490 \\
27   & ~NGC 2158$^{c}$ &  10.0 & 1.00 $^{\pm0.04}$ & 0.35 $^{\pm0.02}$ & 0.79 & 13.8 & 4980 $^{\pm230}$ & 13.46 & 4944& --575 & ~155 \\
28   & NGC 2165 & 10.0 & 1.5 $^{\pm0.06}$ & 0.23 $^{\pm0.05}$ & 0.23 & 11.0 & 1445 $^{\pm67}$ & 9.89  & 1328  & 427    & 377 \\
29   & NGC 2189 & 14.0 & 0.8 $^{\pm0.03}$ & 0.39 $^{\pm0.05}$ & 0.37 & 11.7 & 1869 $^{\pm86}$ & 10.19 & 1641  & --853  & --268 \\
30   & NGC 2220 & 13.0 & 3.0 $^{\pm0.12}$ & 0.06 $^{\pm0.01}$ & 0.06 & 10.4 & 1170 $^{\pm54}$ & 8.92  & 322   & --1020 & --475 \\
31   & NGC 2219 & ~7.0  & 0.8 $^{\pm0.03}$ & 0.40 $^{\pm0.08}$ & 0.41 & 11.9 & 2023 $^{\pm93}$ & 10.24 & 1660  & --1118 & --292 \\
32   & NGC 2224 & 14.0 & 0.01 $^{\pm0.00}$ & 1.00 $^{\pm0.25}$ & 0.99 & 12.8 & 2415 $^{\pm111}$ & 10.81 & 2284  & --785  & 23 \\
33   & NGC 2234 & 28.0 & 0.8 $^{\pm0.03}$ & 0.51 $^{\pm0.10}$ & 0.53 & 11.5 & 1617 $^{\pm75}$ & 10.07  & 1555  & --434  & 79 \\
34   & NGC 2250 & ~4.0  & 0.6 $^{\pm0.02}$ & 0.48 $^{\pm0.10}$ & 0.49 & 11.7 & 1795 $^{\pm83}$ & 10.02 & 1456  & --1031 & --201 \\
35   & NGC 2248 & ~3.0  & 1.0 $^{\pm0.04}$ & 0.23 $^{\pm0.05}$ & 0.22 & 11.4 & 1740 $^{\pm80}$ & 10.23 & 1707  & --226  & 250 \\
36   & NGC 2260 & 20.0 & 0.01 $^{\pm0.00}$ & 1.25 $^{\pm0.20}$ & 1.26 & 12.6 & 1985 $^{\pm90}$ & 10.23 & 1667  & --1071 & --126 \\
37   & NGC 2265 & 12.0 & 0.3 $^{\pm0.01}$ & 0.48 $^{\pm0.10}$ & 0.48 & 12.1  & 2160 $^{\pm100}$ & 10.54 & 2010  & --781  & 123 \\
38   & ~NGC 2266$^{c}$ &  ~5.0 & 0.62 $^{\pm0.02}$ & 0.10 $^{\pm0.05}$ & 0.11 & 12.5 & 3040 $^{\pm140}$ & 11.52& 2963& --405 & ~543 \\
39   & NGC 2312 & ~7.6  & 0.33 $^{\pm0.01}$ & 0.16 $^{\pm0.05}$ & 0.15 & 11.9  & 2245 $^{\pm103}$ & 10.58 & 2030  & --928  & 246 \\
40   & NGC 2318 & 20.0 & 0.05 $^{\pm0.00}$ & 0.65 $^{\pm0.05}$ & 0.65 & 11.2 & 1335 $^{\pm62}$ & 9.48  & 924   & --958  & --104 \\
41   & NGC 2348 & 10.0 & 1.8 $^{\pm0.07}$ & 0.13 $^{\pm0.05}$ & 0.12 & 10.2 & 1070 $^{\pm48}$ & 8.42  & --139 & --969  & --432 \\
42   & NGC 2331 & 14.0 & 1.7 $^{\pm0.12}$ & 0.06 $^{\pm0.02}$ & 0.06 & 10.6 & 1285 $^{\pm59}$ & 9.77  & 1222  & --210  & 337 \\
43   & NGC 2338 & ~7.0  & 0.55 $^{\pm0.02}$ & 0.48 $^{\pm0.05}$ & 0.47 & 11.7 & 1800 $^{\pm83}$ & 9.95  & 1381  & --1154 & 32 \\
44   & NGC 2349 & 18.0 & 0.75 $^{\pm0.03}$ & 0.61 $^{\pm0.10}$ & 0.59 & 11.6 & 1628 $^{\pm75}$ & 9.76  & 1195  & --1106 & 10 \\
45   & NGC 2352 & ~6.0  & 0.12 $^{\pm0.00}$ & 0.32 $^{\pm0.15}$ & 0.31 & 11.5 & 1750 $^{\pm81}$ & 9.57  & 953   & --1455 & --191 \\
\hline
\end{tabular}
\end{center}
\end{table}

\begin{table}
\begin{center}
%\concaption{}
\scriptsize
%\centerline{
\centering
\setlength{\tabcolsep}{0.6mm}
\renewcommand\arraystretch{1.2}
\begin{tabular}{ccccccccccccc}
& & & & & {\bf Table 3.} --- {Continued} & & & & & \\ \\
\hline \hline
Index   & Cluster  & Diam.  & Age  & E$_{B-V}$  & Sch.  & m-M  & Dist.  & R$_{gc}$ & X$_{\odot}$  & Y$_{\odot}$  & Z$_{\odot}$\\
            &   & \it arcmin & \it Gyr & \it mag  & \it mag  & \it mag  & \it pc  & \it kpc   & \it pc & \it pc  & \it pc\\
\hline
46   & NGC 2351 & 10.0 & 0.24 $^{\pm0.01}$ & 0.92 $^{\pm0.25}$ & 0.94 & 12.2  & 1882 $^{\pm87}$ & 9.93 & 1336  & --1325 & 2 \\
47   & NGC 2364 & 13.0 & 0.2 $^{\pm0.01}$ & 0.31 $^{\pm0.05}$ & 0.32 & 11.7 & 1919 $^{\pm88}$ & 10.00 & 1401  & --1307 & 101 \\
48   & NGC 2408 & 28.0 & 3.0 $^{\pm0.12}$ & 0.03 $^{\pm0.02}$ & 0.03 & 10.3 & 1133 $^{\pm52}$ & 9.44  & 794   & 585    & 557 \\
49   & NGC 2455 & ~8.0 & 0.18 $^{\pm0.01}$ & 0.54 $^{\pm0.10}$ & 0.56 & 12.6 & 2650 $^{\pm122}$ & 10.14 & 1389  & --2254 & 107 \\
50   & NGC 2459 & ~5.4  & 1.6 $^{\pm0.11}$ & 0.03 $^{\pm0.01}$ & 0.02 & 10.6 & 1300 $^{\pm60}$ & 9.64  & 1060  & --640  & 397 \\
51   & ~NGC 2588$^{c}$ &  ~2.2 & 0.45 $^{\pm0.01}$ & 0.45 $^{\pm0.10}$ & 0.45 & 13.6 & 4365 $^{\pm201}$ & 10.67 & 1327& --4154& ~186 \\
52   & NGC 2587 & ~4.0  & 0.1 $^{\pm0.00}$ & 0.23 $^{\pm0.10}$ & 0.24 & 11.4 & 1740 $^{\pm80}$ & 9.26  & 609   & --1624 & 136 \\
53   & NGC 2609 & ~5.0  & 0.8 $^{\pm0.03}$ & 0.23 $^{\pm0.10}$ & 0.22 & 10.8 & 1320 $^{\pm61}$ & 8.46  & --138 & --1280 & --291 \\
54   & NGC 2666 & 11.0 & 3.2 $^{\pm0.13}$ & 0.03 $^{\pm0.01}$ & 0.03 & 9.70 & ~860 $^{\pm40}$ & 9.36  & 664   & 47     & 544 \\
55   & NGC 2678 & 13.0 & 2.3 $^{\pm0.09}$ & 0.03 $^{\pm0.01}$ & 0.03 & 9.80 & ~900 $^{\pm41}$ & 9.24  & 621   & --452  & 469 \\
56   & NGC 2932 & 21.0 & 0.5 $^{\pm0.02}$ & 0.55 $^{\pm0.10}$ & 0.56 & 11.4 & 1525 $^{\pm70}$ & 8.59  & --47  & --1521 & 103 \\
57   & NGC 2995 & ~7.0  & 0.05 $^{\pm0.00}$ & 1.94 $^{\pm0.30}$ & 1.94 & 9.60 & ~380 $^{\pm17}$ & 8.46  & --53  & --376  & --8 \\
58   & NGC 3231 & ~7.0  & 1.4 $^{\pm0.06}$ & 0.02 $^{\pm0.02}$ & 0.02 & 9.30 & ~715 $^{\pm33}$ & 9.07   & 401   & 314    & 502 \\
59   & NGC 3446 & 15.0 & 1.0 $^{\pm0.04}$ & 0.16 $^{\pm0.05}$ & 0.15 & 11.0 & 1485 $^{\pm68}$ & 8.32  & --298 & --1417 & 329 \\
60   & ~NGC 3496$^{c}$ &  10.0 & 0.35 $^{\pm0.01}$ & 0.48 $^{\pm0.15}$ & 2.43 $\dag$ & 11.5 & 1640 $^{\pm75}$ & ~8.10 & --548& --1546 & --12 \\
61   & NGC 3520 & ~3.0  & 3.2 $^{\pm0.13}$ & 0.03 $^{\pm0.01}$  & 0.04 & 10.5 & 1245 $^{\pm57}$ & 8.57  & --13  & --978  & 771 \\
62   & NGC 3909 & 16.0 & 2.0 $^{\pm0.08}$ & 0.13 $^{\pm0.05}$ & 0.12 & 10.3 & 1100 $^{\pm50}$ & 8.14  & --409 & --989  & 254 \\
63   & NGC 4230 & 10.0 & 1.7 $^{\pm0.12}$ & 0.23 $^{\pm0.10}$ & 0.24 & 11.0 & 1445 $^{\pm67}$ & 7.92  & --673 & --1265 & 187 \\
64   & NGC 5155 & 17.0 & 1.5 $^{\pm0.18}$ & 0.06 $^{\pm0.02}$ & 2.61 $\dag$ & 10.2 $^{\pm0.1~\Downarrow}$ & 1070 $^{\pm49}$ & 7.9 & --647& --852 & --16 \\
65   & NGC 5269 & ~3.0  & 0.16 $^{\pm0.11}$ & 0.52 $^{\pm0.10}$ & 9.13 $\dag$ & 11.2 & 1410 $^{\pm65}$ & 7.69  & --886 & --1096 & --16 \\
66   & NGC 5299 & 33.0 & 2.0 $^{\pm0.08}$ & 0.19 $^{\pm0.05}$ & 1.94 $\dag$ & 10.4 & 1111 $^{\pm50}$ & 7.83 & --717 & --847  & 40 \\
67   & NGC 5381 & 11.0 & 1.6 $^{\pm0.11}$ & 0.06 $^{\pm0.02}$ & 1.60 $\dag$ & 10.4 & 1170 $^{\pm54}$ & 7.77 & --776 & --874 & 43 \\
68   & NGC 5800 & 12.0 & 0.9 $^{\pm0.04}$ & 0.62 $^{\pm0.10}$ & 0.61 & 12.2 & 2146 $^{\pm99}$ & 6.92  & --1692 & --1301 & 222 \\
69   & NGC 5925 & 24.0 & 0.25 $^{\pm0.01}$ & 0.58 $^{\pm0.10}$ & 2.75 $\dag$ & 10.6 & 1040 $^{\pm48}$ & 7.68 & --845& --606 & 31 \\
70   & NGC 5998 & ~9.0 & 2.2 $^{\pm0.09}$ & 0.16 $^{\pm0.05}$ & 0.15 & 10.1 & ~981 $^{\pm45}$ & 7.56  & --886 & --257  & 332 \\
71   & ~NGC 6005$^{c}$ &  ~6.0 & 1.0 $^{\pm0.04}$ & 0.45 $^{\pm0.20}$ & 0.84 & 12.3 & 2400 $^{\pm110}$ & ~6.65& --1982& --1348 & --125 \\
72   & NGC 6334 & 31.0 & 0.5 $^{\pm0.02}$ & 1.06 $^{\pm0.25}$ & 27.11 $\dag$ & 11.0 & 1025 $^{\pm47}$ & 7.49  & --1013& --158& 8 \\
73   & NGC 6360 & ~5.0  & 0.02 $^{\pm0.00}$ & 1.11 $^{\pm0.20}$ & 1.10 & 11.6 & 1337 $^{\pm62}$ & 7.17 & --1333& --76   & 73 \\
74   & NGC 6357 & ~5.0  & 0.4 $^{\pm0.02}$ & 1.35 $^{\pm0.30}$ & 52.66 $\dag$ & 11.6 & 1205 $^{\pm55}$ & 7.3 & --1196& --143 & 19 \\
75   & NGC 6374 & ~3.6  & 1.3 $^{\pm0.05}$ & 0.48 $^{\pm0.05}$ & 12.65 $\dag$ & 10.2 & ~900 $^{\pm41}$ & 7.6  & --897 & --73  & 7 \\
76   & NGC 6421 & ~8.0  & 0.17 $^{\pm0.01}$ & 1.26 $^{\pm0.20}$ & 1.27 & 12.0 & 1505 $^{\pm69}$ & 7.0   & --1500& --107  & --63 \\
77   & NGC 6437 & 15.0 & 0.2 $^{\pm0.01}$ & 0.71 $^{\pm0.05}$ & 0.70 & 10.5 & ~943 $^{\pm43}$ & 7.56  & --936 & --91   & --69 \\
78   & ~NGC 6451$^{c}$ &  ~7.2 & 0.14 $^{\pm0.01}$ & 0.71 $^{\pm0.05}$ & 1.52 $\dag$ & 12.2 & 2060 $^{\pm95}$ & ~6.44 & -2059& ~--19 & ~--58 \\
79   & NGC 6525 & 13.0 & 2.0 $^{\pm0.08}$ & 0.14 $^{\pm0.03}$ & 0.14 & 10.9 & 1436 $^{\pm66}$ & 7.46  & --1097 & 838  & 393 \\
80   & NGC 6573 & ~1.8  & 0.01 $^{\pm0.00}$ & 2.48 $^{\pm0.20}$ & 2.53 & 10.5 & ~460 $^{\pm21}$ & 8.05  & --454 & 72     & --17 \\
81   & NGC 6595 & ~4.0  & 0.45 $^{\pm0.02}$ & 0.94 $^{\pm0.10}$ & 11.86 $\dag$ & 11.9 & 1640 $^{\pm76}$ & 6.90 & --1607 & 325 & --49 \\
82   & NGC 6605 & 17.0 & 0.6 $^{\pm0.02}$ & 0.52 $^{\pm0.10}$ & 7.99 $\dag$ & 10.2 & ~889 $^{\pm40}$ & 7.65  & --855 & 244 & 5 \\
83   & ~NGC 6603$^{c}$ &  ~6.6 & 0.20 $^{\pm0.08}$ & 0.77 $^{\pm0.10}$ & 3.43 $\dag$ & 13.4 & 3495 $^{\pm160}$ & ~5.15 & --3406& ~778  & --80 \\
84   & NGC 6588 & ~5.0  & 1.6 $^{\pm0.11}$ & 0.10 $^{\pm0.03}$ & 0.08 & 10.0 & ~960 $^{\pm44}$ & 7.68  & --783 & --437  & --342 \\
85   & NGC 6659 & 14.0 & 4.0 $^{\pm0.16}$ & 0.10 $^{\pm0.03}$ & 0.11 & 10.4 & 1155 $^{\pm53}$ & 7.85  & --682 & 888    & 282 \\
86   & NGC 6698 & 11.0 & 1.9 $^{\pm0.07}$ & 0.32 $^{\pm0.05}$ & 0.31 & 10.59& 1150 $^{\pm53}$ & 7.37  & --1115& 182    & --215 \\
87   & NGC 6724 & ~6.0  & 0.9 $^{\pm0.03}$ & 1.00 $^{\pm0.10}$ & 1.00 & 11.1 & 1105 $^{\pm51}$ & 7.73  & --809 & 750    & 69 \\
88   & NGC 6735 & 12.0 & 0.5 $^{\pm0.02}$ & 0.87 $^{\pm0.15}$ & 1.66 $\dag$ & 11.6 & 1466 $^{\pm68}$ & 7.34 & --1209& 828 & --47 \\
89   & NGC 6743 & ~7.0  & 1.4 $^{\pm0.05}$ & 0.19 $^{\pm0.05}$ & 0.18 & 10.4 & 1111 $^{\pm51}$ & 8.01  & --539 & 948    & 211 \\
90  & ~NGC 6755$^{c}$ &  12.8 & 0.05 $^{\pm0.00}$ & 0.84 $^{\pm0.10}$ & 2.19 $\dag$ & 12.0 & 1785 $^{\pm82}$ & ~7.19 & --1394& 1113  & --52 \\
91  & NGC 6773 & ~8.6  & 0.1 $^{\pm0.00}$ & 1.16 $^{\pm0.20}$ & 1.24 & 12.7 & 2160 $^{\pm100}$ & 6.98  & --1653& 1386   & --113 \\
92  & NGC 6775 & ~2.4  & 0.9 $^{\pm0.03}$ & 0.48 $^{\pm0.05}$ & 0.49 & 10.8 & 1185 $^{\pm55}$ & 7.58  & --947 & 705    & --108 \\
93  & NGC 6795 & ~8.0  & 0.95 $^{\pm0.04}$ & 0.45 $^{\pm0.05}$ & 0.45 & 11.0 & 1320 $^{\pm61}$ & 7.54  & --1004& 845    & --141 \\
94  & NGC 6815 & 30.0 & 0.15 $^{\pm0.01}$ & 1.10 $^{\pm0.20}$ & 1.83 $\dag$ & 12.5 & 2024 $^{\pm93}$ & 7.76 & --945 & 1788 & 72 \\
95  & NGC 6832 & 24.0 & 3.0 $^{\pm0.12}$ & 0.10 $^{\pm0.02}$ & 0.09 & 11.3 & 1750 $^{\pm81}$ & 8.74  & 59    & 1678    & 493 \\
\hline
\end{tabular}
\end{center}
\end{table}

\begin{table}
\begin{center}
%\concaption{}
\scriptsize
%\centerline{
\centering
\setlength{\tabcolsep}{0.6mm}
\renewcommand\arraystretch{1.2}
\begin{tabular}{ccccccccccccc}
& & & & & {\bf Table 3.} --- {Continued} & & & & & \\ \\
\hline \hline
Index   & Cluster  & Diam.  & Age  & E$_{B-V}$  & Sch.  & m-M  & Dist.  & R$_{gc}$ & X$_{\odot}$  & Y$_{\odot}$  & Z$_{\odot}$\\
            &   & \it arcmin & \it Gyr & \it mag  & \it mag  & \it mag  & \it pc  & \it kpc   & \it pc & \it pc  & \it pc\\
\hline
96  & NGC 6837 & ~4.6  & 1.0 $^{\pm0.04}$ & 0.25 $^{\pm0.02}$ & 0.25 & 10.1 & ~943 $^{\pm43}$ & 7.93  & --594 & 721    & --131 \\
97  & NGC 6839 & ~6.0 & 1.4 $^{\pm0.05}$ & 0.29 $^{\pm0.05}$ & 0.29 & 11.0 & 1410 $^{\pm65}$ & 7.8   & --783 & 1166   & --127 \\
98  & NGC 6840 & ~6.0 & 1.3 $^{\pm0.04}$ & 0.25 $^{\pm0.05}$ & 0.27 & 11.7 & 1970 $^{\pm90}$ & 7.42  & --1223& 1518   & --283 \\
99  & NGC 6843 & ~5.0 & 1.3 $^{\pm0.04}$  & 0.30 $^{\pm0.05}$ & 0.29 & 11.7 & 1945 $^{\pm90}$ & 7.44  & --1203& 1501   & --284 \\
100  & NGC 6846 & ~4.8 & 0.55 $^{\pm0.02}$ & 0.68 $^{\pm0.05}$ & 2.07 $\dag$ & 11.4 & 1445 $^{\pm67}$ & 8.09 & --525 & 1345 & 48 \\
101  & NGC 6847 & 20.0 & 0.5 $^{\pm0.02}$ & 0.58 $^{\pm0.05}$ & 3.09 $\dag$ & 11.9 & 1894 $^{\pm87}$ & 7.95 & --743 & 1742 & 26 \\
102  & NGC 6856 & ~3.2 & 1.8 $^{\pm0.06}$ & 0.16 $^{\pm0.02}$ & 0.15 & 11.3 & 1704 $^{\pm79}$ & 8.66  & --9   & 1657   & 398 \\
103  & NGC 6858 & 10.0 & 2.5 $^{\pm0.10}$ & 0.13 $^{\pm0.02}$ & 0.14 & 10.7 & 1310 $^{\pm60}$ & 7.75  & --805 & 1007   & --234 \\
104  & NGC 6859 & 10.0 & 3.0 $^{\pm0.12}$ & 0.19 $^{\pm0.05}$ & 0.18 & 10.8 & 1335 $^{\pm62}$ & 7.56  & --957 & 856    & --364 \\
105  & ~NGC 6871$^{c}$ &  32.0 &  0.01 $^{\pm0.00}$ & 0.45 $^{\pm0.05}$ & 2.01 $\dag$ & 11.4 & 1585 $^{\pm73}$ & ~8.17 & --472 & 1512 & ~~57 \\
106  & NGC 6873 & 15.0 & 0.88 $^{\pm0.04}$ & 0.35 $^{\pm0.05}$ & 0.36 & 10.8 & 1250 $^{\pm58}$ & 7.96  & --613 & 1081   & --134 \\
107  & NGC 6895 & 16.0 & 1.0 $^{\pm0.04}$ & 0.35 $^{\pm0.05}$ & 0.34 & 10.6 & 1141 $^{\pm53}$ & 8.5   & --81  & 1126   & 164 \\
108  & NGC 6904 & ~8.0 & 1.0 $^{\pm0.04}$ & 0.39 $^{\pm0.05}$ & 0.40 & 11.0 & 1355 $^{\pm62}$ & 8.05  & --545 & 1232  & --149 \\
109  & NGC 6938 & ~7.2 & 1.3 $^{\pm0.04}$ & 0.13 $^{\pm0.05}$ & 0.12 & 10.6 & 1250 $^{\pm58}$ & 8.05  & --521 & 1112  & --233 \\
110  & NGC 6950 & 15.0 & 1.8 $^{\pm0.05}$ & 0.06 $^{\pm0.02}$ & 0.08 & 10.2 & 1070 $^{\pm49}$ & 8.04  & --499 & 904   & --281 \\
111  & NGC 7023 & 14.4 & 0.12 $^{\pm0.00}$ & 1.10 $^{\pm0.10}$ & 12.02 $\dag$ & 9.70 & ~560 $^{\pm26}$ & 8.65  & 132 & 527  & 137 \\
112  & NGC 7011 & ~4.4 & 0.4 $^{\pm0.01}$ & 1.08 $^{\pm0.10}$ & 3.17 $\dag$ & 11.4 & 1236 $^{\pm57}$ & 8.55  & --40  & 1235   & 13 \\
113  & NGC 7005 & ~4.0 & 2.5 $^{\pm0.10}$ & 0.03 $^{\pm0.01}$ & 0.04 & 10.1 & 1033 $^{\pm48}$ & 7.69  & --689 & 497    & --588 \\
114  & NGC 7024 & ~5.0 & 0.5 $^{\pm0.02}$ & 1.10 $^{\pm0.10}$ & 1.06  & 12.2 & 1760 $^{\pm81}$ & 8.51  & --175 & 1747   & --119 \\
115  & ~NGC 7039$^{c}$ & 11.6 & 0.075 $^{\pm0.00}$ & 0.13 $^{\pm0.05}$ & 0.87 & 10.0 & ~950 $^{\pm44}$ & ~8.52 & ~-35 & ~949 & ~-28 \\
116  & NGC 7037 & ~6.0 & 2.1 $^{\pm0.08}$ & 0.16 $^{\pm0.05}$ & 0.15  & 11.0 & 1485 $^{\pm68}$ & 8.35  & --276 & 1437   & --252 \\
117  & NGC 7050 & ~7.0 & 2.0 $^{\pm0.08}$ & 0.16 $^{\pm0.05}$ & 0.16  & 10.5 & 1179 $^{\pm54}$ & 8.41  & --171 & 1152   & --180 \\
118  & NGC 7055 & ~5.0 & 0.8 $^{\pm0.03}$ & 1.10 $^{\pm0.10}$ & 1.10  & 11.5 & 1275 $^{\pm59}$ & 8.76  & 165   & 1258   & 124 \\
119  & NGC 7071 & ~8.0 & 0.3 $^{\pm0.01}$ & 1.14 $^{\pm0.20}$ & 1.12  & 12.1 & 1684 $^{\pm78}$ & 8.71  & 42    & 1682   & --59 \\
120  & NGC 7084 & 16.0& 1.5 $^{\pm0.06}$ & 0.10 $^{\pm0.05}$ &
0.11  & 9.50 & ~765 $^{\pm35}$ & 8.27  & --239 & 655    & --315 \\
121  & NGC 7093 & 13.0& 0.9 $^{\pm0.04}$ & 0.61 $^{\pm0.05}$ & 0.58  & 11.8 & 1785 $^{\pm82}$ & 8.72  & 32    & 1780   & --135 \\
122  & NGC 7129 & ~7.0 & 0.12 $^{\pm0.01}$ & 0.97 $^{\pm0.05}$ & 7.01 $\dag$ & 11.0 & 1070 $^{\pm49}$ & 8.84 & 280 & 1016 & 184 \\
123  & NGC 7127 & ~5.0 & 0.4 $^{\pm0.02}$ & 0.90 $^{\pm0.05}$ & 2.17 $\dag$ & 11.6 & 1445 $^{\pm67}$ & 8.82  & 199 & 1431 & 29 \\
124  & NGC 7134 & ~3.0 & 3.3 $^{\pm0.13}$ & 0.06 $^{\pm0.02}$ & 0.05  & 10.2 & 1065 $^{\pm49}$ & 7.74  & --558 & 502    & --755 \\
125  & NGC 7175 & 32.0& 0.25 $^{\pm0.01}$ & 0.87 $^{\pm0.05}$ & 1.72 $\dag$ & 12.2 & 1930 $^{\pm89}$ & 9.03 & 326 & 1902 & --3 \\
126  & NGC 7193 & 13.0& 4.5 $^{\pm0.18}$ & 0.03 $^{\pm0.01}$ & 0.05  & 10.2 & 1080 $^{\pm50}$ & 8.2 & --304 & 839    & --608 \\
127  & NGC 7352 & ~9.0 & 0.05 $^{\pm0.00}$ & 1.10 $^{\pm0.20}$ & 1.08  & 13.0 & 2550 $^{\pm117}$ & 9.52  & 698   & 2452   & --47 \\
128  & ~NGC 7380$^{c}$ &  13.0 & 0.014 $^{\pm0.00}$ & 0.61 $^{\pm0.05}$ & 9.43 $\dag$ & 12.2 & 2145 $^{\pm99}$ & ~9.36 & ~~632& 2049  & ~--33 \\
129  & NGC 7394 & ~9.0 & 0.6 $^{\pm0.02}$ & 0.35 $^{\pm0.05}$ & 0.34  & 10.9 & 1310 $^{\pm60}$ & 8.92  & 332   & 1259   & --147 \\
130  & NGC 7429 & 14.0 & 0.04 $^{\pm0.00}$ & 1.16 $^{\pm0.10}$ & 2.34 $\dag$ & 11.4 & 1190 $^{\pm55}$ & 8.96  & 387 & 1125 & 6 \\
131  & NGC 7686 & 16.0 & 2.0 $^{\pm0.08}$ & 0.20 $^{\pm0.05}$ & 0.19  & 11.1 & 1534 $^{\pm71}$ & 9.13  & 502   & 1416   & --307 \\
132  & NGC 7708 & 24.0 & 2.0 $^{\pm0.08}$ & 0.42 $^{\pm0.05}$ & 0.43  & 11.4 & 1607 $^{\pm74}$ & 9.35  & 726   & 1401   & 301 \\
133  & NGC 7795 & 21.0 & 0.45 $^{\pm0.02}$ & 1.00 $^{\pm0.10}$ & 1.03  & 12.5 & 2105 $^{\pm97}$ & 9.62  & 935   & 1885   & --79 \\
\hline
\end{tabular}
\end{center}
\scriptsize $^{\rm c}$ Calibrated clusters, which are listed in Table 2.\\
\scriptsize $^{\rm \dag}$ Unreliable/Overestimated reddening values.
\end{table}

%--------------------------------------------------------------------------------------
\end{document}